\documentstyle[11pt,aaspp4]{article}
\begin{document}

\title{Cosmological Evolution Models for QSO/AGN Luminosity Functions:
Effects of Spectrum-Luminosity Correlation and Massive Black Hole Remnants}
\author{Yun-Young Choi$^{1,2}$, Jongmann Yang$^{1,2}$, and Insu Yi$^{3}$}
\affil{$^1$Center for High Energy Astrophysics and Isotope Studies,
Research Institute for Basic Sciences}
\affil{$^2$Department of Physics, Ewha University, Seoul 120-750, Korea}
\affil{$^3$Korea Institute for Advanced Study, Seoul 130-012, Korea}

\begin{abstract}
We investigate a large number of cosmological evolution models for
QSOs and Active Galactic Nuclei (AGN). We introduce a
spectrum-luminosity correlation as a new input parameter and adopt
the estimated mass function (MF) of massive black holes in centers
of nearby galactic nuclei as a constraint to distinguish among
different QSO/AGN models. We explore three basic types of
phenomenological scenarios; (i) Models with multiple short-lived
($\sim$ a few $10^{6-8}$ yrs) populations, (ii) Models with a
single long-lived ($\sim 10^9$ yr) QSO population, and (iii) Models
with recurrent QSO/AGN activities which are driven by long-term
variabilities of the disk instability type. In each model, we
derive the expected theoretical luminosity function (LF) and the MF
of black holes which grow through mass accretion. We assess the
plausibility of each model based on whether each model's LF and MF
are compatible with the observed data. We find that the best fits
to the observed LFs are obtained in the model with multiple
short-lived populations and without any significant spectral
evolution. This finding suggests that the QSO populations may be
composed of many short-lived generations ($\sim$ a few 10$^8$ yrs) and
that there is no significant spectral evolution within each
generation. On the other hand, we also show that there is no
satisfactory model which can simultaneously account for the
observed LF and the estimated MF. We speculate that some of the
present-day black holes (BHs) found in galactic nuclei may have
formed without undergoing the QSO/AGN phase.
\end{abstract}

\section{INTRODUCTION}

It has been well established that the observed comoving space density of
optically bright QSOs reaches a peak at a critical redshift $z\sim 2$
(e.g. Peterson 1997, Hartwick \& Shade 1990, Weedman 1986).
A similar trend is also seen in the X-ray evolution
(Miyaji et al. 2000 and references therein). In both cases, the
evolution of the luminosity function (LF) is roughly accounted for by
the number-conserving luminosity evolution (e.g. Mathez et al. 1976)
in which the luminosities of
QSOs first gradually decrease at $z>2$ from their births near $z>4$ and
rapidly decline at $z<2$. It has recently been questioned whether the
X-ray evolution can be adequately described by the pure luminosity
evolution (Hasinger 1998, Miyaji et al. 2000).
Although the critical redshift at which the QSO activities
show a suddden transition is firmly established, it is unclear what
determines such a redshift. There have been numerous attempts and models
leading to debates as to what physical processes in QSO/AGN engines and/or
their surroundings determine the basic characteristics of the cosmological
evolution of QSO/AGN (e.g. Caditz et al. 1991, Small \& Blandford 1992,
Fukugita \& Turner 1995, Yi 1996, Haehnelt et al. 1998, Nulsen \& Fabian 2000).

We intend to explore most of the existing classes of QSO/AGN evolution
in their broad categories. The phenomenological scenarios we explore in
this work can be roughly classified as follows. First, we consider a
class of models in which a single long-lived ($\geq 10^9$ yr) QSO
population evolves throughout the cosmological time after birth
at high redshifts $z>4$ (Mathez 1976, Yi 1996, Peterson 1997 and
references therein).
Second, we study the models in which many
short-lived (a few $10^8$ yr) QSO populations form and evolve. In these
models, the overall observed evolutionary trend is a result of the
collective evolution of many different generations of QSO/AGN
(Haehnelt et al. 1998). Third,
a model of recurrent QSO/AGN activities in which some long-term
variabilities of QSO/AGN emission dominates their recurrent activities
(Siemiginowska \& Elvis 1997).
As a specific model of the third type, we assume that the variabilities
are mainly caused by the accretion disk instability of the dwarf novae
type (Frank et al. 1992) as is occasionally discussed in connection with
the origin of high luminosity QSO activities.

We consider two fundamentally different assumptions which are motivated by the
recent works in the general area of acrretion disk physics. These
assumptions have direct implications on the luminosity evolution of
the QSO/AGN and as a consequence they are indirectly testable.

The first one of the two is that the change of accretion flow around a
central black hole, which powers the QSO/AGN activities, is mainly driven
by the change in mass accretion rate. Such a change then results in
the correlated spectral and luminosity evolution of QSO. This assumption
is based on the observed behavior of black hole X-ray binaries (BHXBs).
Some BHXBs show prominent spectral changes which are strongly correlated
with the luminosity changes (e.g. Rutledge et al. 1999, Choi et al. 1999a, 1999b, 2000).
Although the details of the models accounting
for this behavior vary significantly, it is widely believed that the changes
in the mass accretion rate is the underlying cause of this phenomenon.
For instance, the ADAF models (Narayan et al. 1998, Yi 1996 and references
therein) have enjoyed a relative success in explaining
the hard X-ray emitting states of BHXBs while the thin disk models
(Frank et al. 1992) have been
widely applied to the soft X-ray emitting states. In these models, the
continuous changes in the spectral state is interpreted as changes in
the accretion rate and the accretion disk's physical conditions
(e.g. Esin, McClintock, \& Narayan 1997, and reference therein,
Rutledge et al. 1999). We assume that the QSOs' luminosity evolution is
accompanied by a strong spectral evolution caused by the accretion flow
transition (see Choi et al. 1999a for details).
The spectral state is determined by the BH mass and the physical
accretion rate. The dominating effect is obviously that the resulting
luminosity evolution shows distinct evolutionary behavior in
different energy bands. The nearly mass scale-invariant nature of the
accretion flow properties further support this assumption (Narayan \& Yi
1995).

The second one is that the QSOs luminosities are interpreted as fixed
fractions of the bolometric luminosity and each band's luminosity is simply
given as a fixed fraction of the bolometric luminosity regardless of changes
in the accretion rate. The latter approach is closer to those of the
conventional studies carried out so far (Peterson 1997).
Although this assumption is not
clearly supported by any physical models of accretion flows, the lack of
detailed QSO spectral information in different bands makes it hard to rule
out this simple but convenient approach.

In our previous work (see Choi, Yang, \& Yi 1999b, 2000 for details), we looked
into the pure luminosity evolution model (Mathez 1976, Peterson 1997)
with the explicit inclusion of
the spectrum-luminosity correlation. In this model, all QSOs are long-lived
and they become gradually dimmer with their comoving space number density
conserved throughout the evolution, i.e. from their roughly synchronous
births to the present epoch. We arrived at a conclusion that the
first assumption, i.e. the accretion flow transition, can be accommodated
in the QSO evolution with relatively good fits to the observed luminosity
evolution. According to this model, however, the smaller mass black holes
with the masses $< 10^8 M_{\odot}$ in galactic nuclei
(Magorrian et al. 1998) cannot be direct
remnants of the past QSO activities as the QSOs' black holes grow much
more massive than these black holes (cf. Wandel 1999).
In this model, the QSO remnants have
to exist in the nuclei of rare massive galaxies. Then, the often discussed
massive dark objects in ordinary galaxies (Magorrian et al. 1998)
should have formed and grown
without experiencing the QSO phase, which does not appear to be a popular
proposition.

In the present work, we explore a number of QSO/AGN evolution models and make
both qualitative and quantitative comparisons among them. The three types
of the models we consider are (i) the single long-lived population,
pure luminosity evolution model (Yi 1996, Choi et al. 1999b),
(ii) the density evolution model in which the QSOs' evolution
is the superposition of multiple generations of short-lived
($\sim$ a few $10^6-10^8$ yr) QSOs (Haehnelt et al. 1998),
and (iii) the model in which the QSO
luminosities undergo recurrent variabilities (Siemiginowska \& Elvis 1997).
The second type of the models have been studied recently
(e.g. Haehnelt \& Rees 1993, Haehnelt, Natarajan, \& Rees 1998) and
they are specifically based on the idea that the hierarchical build-up
of normal galaxies and the evolution of the AGNs/QSOs are closely connected
to each other. It has been claimed that such models are
supported by the recent observational evidences such as
the existence of massive black holes (BHs) with $ 10^{6-10} M_{\odot}$
(e.g. Franceschini et al. 1998), and the strong correlation in their masses
between the supermassive BH and the spheroidal components of nearby galaxies
(e.g. Magorrian et al. 1998, cf. Wandel 1999).
In addition, we also explore a specific physical model in which the QSOs'
long-term variabilities caused by accretion disk instabilities contribute
considerably to the observed QSO LFs (Siemiginowska \& Elvis 1997). We use
these classes of models and make comparisons among them using the
derived LFs and MFs along with the observational data.

In sections 2 and 3, we summarize the evolution models and describe how we
determine model parameters in each model. We derive the resulting analytical
LFs of QSO/AGN and MFs of BH remnants. In section 4, we draw our conclusions
and discuss their implications on how to interpret the observational data.

\section{The Multiple Population Model for Short-Lived QSOs}

In this type of scenario, the formation and evolution of normal galaxies occur
within a hierarchical merging of dark matter halos
(Peebles 1993), which is closely connected
to the formation and evolution of QSOs experiencing short active phases
(Haehnelt \& Rees 1993, Kauffmann \& Haehnelt 2000, Monaco et al. 2000).
It has been generally assumed that a QSO is born (e.g. Nulsen \& Fabian 2000
and references therein)
or re-activated (e.g. Small \& Blandford 1992) when
two galaxies merge and that mergers provide the fuel for the newly
formed central massive BH. It remains unclear whether during the short,
active emission phase, the QSOs' emission spectra show any rapid spectral
evolution while their luminosities rise and fall on time scales much shorter
than the cosmological evolution time scale. In other words, one could ask
how the assumed spectral evolution, which we mentioned above, affects
the QSO LFs in this scenario. In the multiple population models,
the cosmologically evolving QSO population is composed of many short-lived
generations and undergoes successive rise and fall of QSO activities.
If the luminosity-spectrum correlation is applied to the QSO generations,
the evolution of an individual short-lived QSO should exhibit
some appreciable changes in the spectral emission state with the decreasing
mass accretion rate or luminosity after merge-driven trigger of the QSO
activity. One of the most significant constraint on this scenario comes from
the observed LFs (e.g. {\it ROSAT} samples by Miyaji et al. 2000)
and the MF derived from the radio LF of E/SO galaxies cores
(e.g. Salucci et al. 1999).
The AGN LFs in the soft X-ray band have been updated by more extensive
analyses with more recent and expanded {\it ROSAT} Bright Survey and
{\it ROSAT} Deep Survey (Miyaji et al. 2000). Together with the
earlier results reported by Hasinger (1998), these LFs show an apparent excess
at the faintest soft X-ray luminosities $<10^{42}$ erg s$^{-1}$ at the redshift
epoch of $z=0.0-0.2$. In this work, we do not consider this excess
in deriving the best fits based on the fact that at such a low luminosity
level the distinction between QSO/AGN and less active bright galaxies
and/or Seyferts is quite obscure
(e.g. Yi \& Boughn 1998, 1999 and references therein).
For these low luminosity AGN, the observed low-energy X-ray
background (XRB) could provide an integral constraint on their LFs
(e.g. Franceschini et al. 1999).

We describe the several assumptions and parameters involved in deriving
the LFs and constructing an evolution model for QSOs.
We first need to specify an individual QSO activity in terms of its spectra
and luminosities (see, Choi, Yang, \& Yi 1999b for details).
We assume that each QSO begins their activity with a newly formed black hole
of mass, M, and gas accretion flow with a rate, ${\dot M}$, which shows a
generally decreases as the QSO evolves. The rate decrease is taken for
simplicity to be of the exponential form with the characteristic e-folding
time scale $t_{evol}$ (e.g. Haiman \& Menou 2000)
\begin{equation}
\dot M =\dot M_o exp\left(-t/t_{evol}\right)
\end{equation}
where $t=0$ is taken to be the QSO trigger time (e.g. Yi 1996).
The $t_{evol}$ is taken as a fraction of the cosmic time
$t_{age} \sim 10^{10}$ yr for a flat universe (i.e. $q_o = 0.5$) with no
cosmological constant and $H_o =50$ km s$^{-1}$ Mpc$^{-1}$ and essentially
corresponds to the lifetime of a QSO.
The accretion time scale $t$ scales with redshift
\begin{equation}
t \propto (1+z)^{-1.5}
\end{equation}
and the accretion rate,
\begin{equation}
{\dot m}={\dot M}/{\dot M_{Edd}} \propto {\dot M}/M
\end{equation}
is expressed in units of the mass-dependent Eddington accretion rate.

\subsection{The Spectral Evolution Model for Multiple QSO Populations (SEM)}

As ${\dot m}$ decreases after a merging or some other QSO trigger event,
QSOs experience two types of spectral transition, from ``Very High''
state (VHS, slim disk, and ${\dot m} > 1$) to ``High''
state (HS, thin disc, and $0.01\leq \dot{m}\leq 1$) and subsequently,
from ``High'' state to ``Low'' state including
``Off'' state (LS/OS, ADAF, and $\dot{m} < 0.01$). These transitions are
caused by the physical changes in the accretion flows  and the accompanied
luminosity change is strongly correlated with the spectral state change
(Yi 1996, Choi et al. 1999a, 1999b, 2000, Narayan et al. 1998).
This essentially describes the main ingredients in the assumption of spectral
evolution.
Based on this simple assumption, we show the expected spectral energy
distributions in various spectral states of a QSO with
a BH mass of $10^{8} M_{\odot}$ which include the emission from ADAFs
(see Figure 1(a)). The hard X-ray emitting ADAF emission has been
pointed out as a possible source of the diffuse X-ray background (Yi \& Boughn
1998, Fabian \& Rees 1995, Di Matteo et al. 1999).
Figure 1(b) shows the expected luminosity evolution of a long-lived QSO
with the evolution time scale
\begin{equation}
t_{evol}=0.50t_{age}\simeq 6.4\times 10^9~yr
\end{equation}
where the initial BH mass is $10^8 M_{\odot}$ and the initial $\dot m=1$.
The different energy bands show different luminosity evolution trends
(Choi et al. 1999b), which
suggest that the spectral evolution could be distinguishable from the simple
non-spectral evolution case in which the luminosity in each band is simply
proportional to the bolometric luminosity
(see a dash-dotted line in Figure 1(b)).
The transition of accretion flows from the thin disk type (HS) to the ADAF
type (LS/OS) occurs at the critical accretion rate
$\dot m_{c}=0.3 \alpha^2$ where $\alpha=0.3$ (Yi 1996, Narayan \& Yi 1995) and
such a transition makes QSO luminosities decrease/increase rapidly at $z<1$.
In particular, the LFs in the hard X-ray band are affected differently
by the spectral change (from HS to LS/OS), which is quite distinct from
the way other energy bands behave. The bolometric luminosity evolution is
primarily determined by the redshift dependence of the evolutionary time scale
defined above. That is, the rapid decline of the QSO's bolometric luminosity
at low redshifts
is caused by the particular redshift dependence of the evolution
time scale we have adopted.

We must also specify the several major parameters necessary for the
construction of the LF in the SEM model. The main physical parameters are:
(i) the merger rate of host galaxies or the birth rate of QSOs, (ii) the life
time of an individual QSO, and (iii) the mass distribution of seed BHs at birth.
The birth rate in this work is equivalent to the formation rate of new seed
BHs at different redshifts. We assume that the birth rate is a simple function
of the redshift for which we adopt a simple functional form.
For simplicity, we adopt the power-law of the form
\begin{equation}
N(z) = N_o [(1+z) / (1+z{_o})]^{s}
\end{equation}
where $N_o$ is the number of QSOs belonging to the first generation at the
epoch, $z=z_o$, and $s$ determines the basic evolutionary trend of
the birth rate as a function of the redshift.

We note that all of the characteristic parameters we find in each evolution
model can cause substantial differences to the QSO number density and LF.
The best fit values of the parameters are determined by
comparing the resulting LF fits in the soft X-ray band with
the observed LFs of AGN in the same band. Simultaneously,
the theoretically predicted mass distributions are compared with the estimated
MF from radio data of E/S0 galaxies. In this sense, the mass distribution of
QSO remnants provides an additional constraint. Such a constraint is of course
invalid if a significant fraction of black holes at galactic nuclei are not
formed from QSO activities.

Figure 2 clearly shows the dependence of the QSO LF on the three main
parameters. In other words, in the SEM models considered, the LFs are
essentially determined by the number density in each QSO generation,
the number of generations, g, constituting QSOs population,
and the evolutionary time scale.
QSO LFs shown in this figure refer to those in the soft X-ray energy band,
which is identical to the observed soft X-ray band. We assume that the initial
accretion rate $\dot m$ of an individual QSO dose not exceed the Eddington
limit, $\dot m=1$.
The comoving number density is normalized to match that of Miyaji et al.
(2000) in panel (b).
Panel (a1) gives the best fit model which is in good agreement
with the observed LFs (see panel (b)). This particular model assumes
with the simple power-law evolution law ($s=-1$) with a broad
initial mass distribution of seed BHs, and 100 QSO generations
over the range of $0.18 \leq z \leq 4.3$.
The $t_{evol}$ for all QSOs is given as $0.03~t_{age}\simeq 3.9\times 10^8$
yr. This is in the lifetime range ($\sim 10^6 - 10^8$ yrs) of the luminous phase of QSOs
suggested by Haiman \& Hui (2000).
The initial masses of seed BHs in each generation are randomly chosen by
a single power law with a slope of $2.5$ and
their masses are spread from $10^{7-10} M_{\odot}$ in
the first high $z$ generation to $10^{6-9} M_{\odot}$ in the last low
$z$ generation. Although these assumptions are rather arbitrary,
the best fit model is clearly contrasted with less successful models
in fitting the observed LF in panel (b).

Panel (c1) shows the predicted MF of BHs at six redshifts
obtained from LFs (in panel (a1)) which are in turn derived
in the SEM model and panel (c2) illustrates
that in the single QSO population model (hereafter SES model and see below).
The solid line with the cross symbols in panel (c1) and (c2)
represents the estimated MF from radio data of E/SO galaxies
in Salucci et al. (1999).
According to this model, just a few \% of the present-day galaxies with the
comoving number density $\sim 10^{-2}$ h${_{50}}^3$Mpc$^{-3}$ (Magorrian
et al. 1998) would have passed through the QSO phase (cf. Wandel 1999).
On the other hand, due to the longevity of the QSOs
and the prolonged accretion phase,
it is difficult for a single populations model to explain the low mass
BHs with masses of $10^{6-7} M_\odot$ in nearby spiral galaxies
(Choi et al. 1999b, 2000 and references therein).
In the single population model, the massive dark objects in the nearby galactic
centers cannot have passed through the long QSO phase (see panel (c2)).
It is still possible that some rare massive black holes in giant elliptical
galaxies residing in galaxy clusters
(Fabian \& Canizares 1988, Mahadevan 1997)
could have grown through a long-lasting
QSO phase. As the number of QSO generations in the SEM model decreases, the
resulting LF become similar to those of the single population model
as anticipated.
The apparent transition from SEM to the single population model
occurs when the number of QSO generations falls to $\sim 10$ and the QSO
evolution time scale $t_{evol}\sim 3.9\times 10^9$ yr (in panel (a4)).

One of the distinguishable conclusions concerning this model is that
the number density of bright QSOs is not a constant but shows a strong evolution
over the cosmological time scale (Miyaji et al. 2000, Wisotzki 2000a).
The evolutionary time scale, $t_{evol}$, and its redshift dependence
play an important role in causing this feature.
In essence, a shorter $t_{evol}$ allows $\dot M$ to decline more rapidly
and the BHs to gain less mass. For instance,
for $t_{evol}=0.03t_{age}\simeq3.9\times10^8$ yr, remnant BH masses are bigger
than the initial masses by a factor $\sim 10$.
As a result, QSOs experience the sudden changes of the spectral states and
a large number of QSOs at any given redshift become too faint
to be observed (i.e. $< 10^{42}$ ergs s$^{-1}$).
This sudden decrease in the number of luminous QSOs
gets more marked at lower redshifts due to the particular redshift dependence
of $t_{evol}$. Therefore, unless the power $s$ of the power-law density
evolution reverses its sign from positive to negative (from $+1$ to $-1$),
the evolution trend of number density in the derived LF cannot match the
observed one. In short, the observed LF evolution requires that the number
of QSOs formed at lower redshifts be higher than that at higher redshifts
(in panel (a1) and (a2)).
The parameters used in panel (a2) except for the power $s$ are as given
in panel (a1). Panel (a3) shows that the derived
LFs do not evolve significantly if QSOs formed in all generations have the same initial
mass distribution in the range of $10^6$ to $10^9 M_{\odot}$.
By comparing panel (a1) with panel (a3), we see that when the QSOs with less
massive seed BHs are born and refueled more abundantly at lower redshifts,
the resulting LFs in the SEM model are generally in better agreement with
the observed LFs.

Panels (a1), (a4), (a5), and (c1) show the effects of the number of
generations and the evolutionary time scale which directly affect
the LF evolution and the MF of the BH remnants.
The parameters except for the $t_{evol}$ and the number of generations
used in panels (a4) and (a5) are identical to those given in panel (a1).
If the number of QSO generations is small, the QSO evolution could show
some discontinuous evolutionary behavior. In this case, the birth and death
of a generation of QSOs could show up as a discrete evolutionary episode.
This type of the discrete evolutionary behavior is smeared out when the number
of QSO generations is large enough or the lifetime of QSOs is long enough.
For instance, when QSOs in each generation evolve with
$t_{evol} \simeq 3.9 \times 10^8$ yr, $100$ generations are enough to ensure
that the QSO evolution appears smooth and continuous.

\subsection{Constraint from Mass Function of Supermassive Black Hole Remnants}

The MF provides an additional constraint on the QSO evolution.
In panel (c1), the solid, dash-dotted, and dashed lines (without symbols)
correspond to the predicted MFs of BH remnants in the models with the
LFs in panel (a1), (a4) and (a5), respectively.
Because QSOs with a short lifetime do not substantially grow in mass
during accretion, in order to match the comoving number density implied
by the BH remnant distribution in nearby galaxies, the shorter QSO lifetime
demands the higher QSO number density.
Panel (c1), however, shows that such a compensation can be ruled out in certain
cases based on the MF. Assuming that the QSOs shine with an X-ray
luminosity efficiency of 10\% of the bolometric luminosity
\begin{equation}
L=\eta\dot M c^2
\end{equation}
with the bolometric radiative efficiency $\eta \simeq 0.1$ during QSO phase,
the predicted MF can not account for the high comoving mass density implied by
the nearby galaxies (Salucci et al. 1999).

We find that the comoving number densities of the BH
remnants inferred from the best-fitted soft X-ray LF (in panel (a1))
are smaller than those estimated for the putative BHs in the bulges
of nearby galaxies (e.g. Magorrian et al. 1998, Salucci
et al. 1999) by a factor of about $100$. We also find that the
same number densities are smaller by a factor of about $10$ than those
implied by optically bright QSOs under the assumption that they
are powered by accretion onto supermassive BHs with an assumed accretion
efficiency of $10\%$ (e.g. Phinney 1997, Haehnelt et al. 1998
and reference therein).
These differences are significant and there do exist some discrepancies
between our estimates and the other earlier estimates.
Most notably, we use the soft X-ray LF to match the MF of BHs while
other studies adopted the B-band LF to match the MF of massive dark
objects.
In order to allow the predicted MFs to satisfy observational constraint,
a number of conditions have to be met.
First, the lifetime of sample QSOs must be short (a few $10^6$ to
$10^7$ yr) and the QSOs have to be already massive enough
(about $10^8$ to $10^{11} M_{\odot}$) before accretion-driven mass growth
occurs. Such massive black holes could form without emitting detectable
photons if their pre-QSO evolution occurs in the form of the super-Eddington
accretion with very low radiative efficiencies (Haehnelt et al. 1998).
Second, the X-ray fraction of the bolometric luminosity,
$f_{XR}$, and/or the bolometric radiative efficiency, $\eta$ must be smaller
than that previously assumed (Haehnelt et al. 1998).
For instance, in the absence of any spectral evolution in the context of
the multiple QSO population model (hereafter NEM model),
the best-fit value of $f_{XR} \cdot \eta$ is smaller than the usually assumed
value by a factor of about 100.

\subsection{No Spectral Evolution Model with Multiple Populations (NEM)}

In Figure 3 (a1), we plot the soft X-ray LF
giving the best-fit MF of BH remnants (dashed line in Figure 3 (c))
for the model of no-spectral evolution of multiple QSO population
(hereafter NEM model). In Figure 3 (a2) and (c),
the best-fit soft X-ray LF and the MF (dash-dotted line)
corresponding to this LF
are shown. The resulting LF in Figure 3 (a1)
is in very poor agreement with the observed LF in Figure 3 (b).
Figure 3 demonstrates the difficulty of reconciling the LFs and MFs and
fitting them simultaneously.

\subsection{Discrepancies between MF and LF}

There are several possibilities which could potentially account for the
discrepancies between the LF and the MF. These possibilities essentially
rely on the formation and accretion history for supermassive BHs.

First, the mass density derived from the Massive Dark Objects (MDOs) mass
estimates (e.g. Magorrian et al. 1998) could be grossly overestimated due to
some systematic uncertainties in the estimates of $L_{sph}/L_{tot}$,
$M_{MDO}/M_{sph}$, etc. (e.g. Salucci et al. 1999).
Second, it might be unreasonable for all galaxies to contain
MDOs which are actually supermassive BHs. In other words,
just a few \% of the present-day galaxies with the
comoving number density $\sim 10^{-2}$ h${_{50}}^3$Mpc$^{-3}$ (Magorrian
et al. 1998) could have passed through the QSO phase powered by
accretion on to supermassive BHs.
Third, the low optical emission efficiency could be responsible for
low number densities observed in the LFs (Haehnelt et al. 1998).
Haehnelt et al. (1998) speculated on two possibilities based on the
possibility of the low optical emission efficiency. One of them is the
existence of a population of obscured QSO by dust based on the unified
Seyfert scheme (e.g. Madau et al. 1994, Comastri et al.
1995, Hasinger 2000, Fabian 1999)
and the other is the existence of a population of galaxies undergoing
ADAF-type accretion (Narayan \& Yi 1995, Fabian \& Rees 1995, Yi 1996).
We emphasize that although
we do not use the B-band LF but the soft X-ray LF to match the observed
soft X-ray LF, our treatment of accretion and QSO emission does include
the evolution through the ADAF phase (via spectral evolution).
Despite this improvement which addresses the low efficiency ADAF flows,
the resulting comoving number density of BH remnants is still far short of
the required mass density of BHs in nearby galaxies.

The gas and dust obscuration could affect the observed QSO luminosities
not only in the optical band but also in the soft X-ray band
(Hasinger 2000, Fabian 1999).
The effects of obscuration in QSO number counting could be significant and its
indirect evidence could be found in the fact that the space density of
obscured AGN is about 3 times higher than that of unobscured AGN in the X-ray
background model (Comastri et al. 1995, Hasinger 1998).
The effects of obscuration on
emission could differ significantly from band to band.
The space density obtained for the soft X-ray of AGN by
Miyaji et al. (2000) exceeds the one observed by Boyle et al.
(1991, 2000), which suggests that the optical and X-ray LFs evolve quite distinctly
(cf. Hatziminaoglou et al. 1998, Wisotzky 2000b).
Therefore, we can plausibly conclude that the effects of the low efficiency
accretion as discussed by Haehnelt et al. (1998) are at most marginally
significant whereas the possibility of obscured accretion and emission
appear to be more important in reconciling the QSO LFs and MFs.

On the other hand, due to the long accretion phase and huge mass gains
of QSOs, it is difficult for a single population model to account for
the smaller mass BHs with masses of $10^{6-7} M_\odot$ in nearby spiral
galaxies in terms of the QSO remnants which have evolved through the
QSO phase (see Figure 2 (c2)). It is also difficult to rule out
the possibility that only the more massive black hole remnants have gone
through the QSO phase and these giant QSO remnants are found in present-day
giant elliptical galaxies. LF in the SEM model with the generations of $10$ and
$t_{evol}$ of $3.9\times 10^9$ yr (in Figure 2 (a4) ) is already nearly
indistinguishable from one in
the single population model with spectral evolution.

\subsection{Comparison among Various Evolution Models}

Figure 4 makes a comparison among the derived LFs in various energy bands
including hard X-ray energies ($2 -10$ keV), soft X-ray energies ($0.5 - 2$ keV), and
$4400 \AA$ for all the QSO evolution models we test.
 The evolution parameters in each model are determined by requiring them to
make the derived the soft X-ray LF in good agreement with
the observed LFs of AGN in the soft X-ray band (Miyaji et al. 2000).
Such a fitting also fixes the overall normalization of the comoving number
density.  The best-fit parameters for various models are shown
in the Table 1.

The MFs of BH remnants in Figure 5 result from these best-fitted
LFs shown in Figure 4. Figure 4 and 5 clearly demonstrate the difficulties
in satisfying the LF and MF constraints simultaneously.
In Figure 4, panel (a1) and (b1) show the derived LFs
in various energy bands for SES and SEM models, respectively.
It appears that the SEM model gives the LFs in the soft X-ray energy band
which are in better agreement with the observed number densities and
luminosities evolution of QSOs than the SES model.
Such an assessment critically relies on the fact that at $z=1.6-2.3$
the number densities of the observed bright QSOs (Miyaji et al. 2000)
apparently decrease only very gradually.

It is generally difficult to interprete specific best-fit values of the QSO
evolution parameters in the multiple population model
although in the hierarchical galaxy formation scenario the QSO evolution occurs
as a part of the general galaxy formation. In particular, the duration of each
QSO generation and the number of QSO generations over the cosmological age of
the universe are not well constrained.
We find that just as in the SES model, the SEM model also gives
different LF evolution trends in different energy bands, which is clearly a
distinguishable and potentially observable feature caused by the strong
spectral evolution correlated with the luminosity evolution.
The optical ($4400 \AA$) LF tends to evolve more rapidly than the X-ray LF at
lower redshifts and the B band LF of QSOs observed by Boyle et al. (1991) in Figure 4.
At lower redshifts, the adopted spectral evolution prescription (Choi et al.
1999b and references therein) causes the QSOs to be optically much less
brighter than the observed QSOs.
The distinctive evolutionary feature at the hard X-ray energies
shows up in the SES model. This feature is essentially summarized as the
the reversal in the direction of luminosity function evolution. Such an
apparently puzzling feature is solely due to the accretion flow transition
from a thin disk to an ADAF. In the SEM model, however, this effect
is cancelled out due to a successive evolution of multiple generations,
a result which doesn't sensitively depend on the exact number of QSO
generations as long as the SEM model has a large enough generations to be
qualified as a multiple population model. It is interesting to point out that
we can infer from the existence or non-existence of this particular feature
whether the QSO population is composed of multiple generations or a single
generation.

\subsection{Multi-Band QSO Luminosity Evolution}

We now look into how the multiple population model is distinguished from the
single population model if the QSO luminosity evolution is not accompanied
by the spectral evolution. In the case of no-spectral evolution, the
luminosity, $L$, in each band is simply
proportional to the bolometric luminosity, $L_{bol}$, i.e.
\begin{equation}
L = f L_{bol} = f (\eta \dot M c^2),
\end{equation}
where $\eta \sim 0.1$ is the radiative efficiency. We adopt the fraction $f$
of each band as 0.43, 0.47, 0.06, and 0.04 in the bands $4400 \AA, 1216 \AA$,
soft X-ray, and hard X-ray, respectively.
These fractions have been chosen solely based on the need to fit the observed
LF evolution in each band and hence at this point,
we don't have any compelling physical motivation for these numbers.
As anticipated, the evolution in all bands
essentially have similar trends.
The fraction of the bolometric luminosity radiated in the optical band of
$4400 \AA$, $f_{4400\AA}$, is about four times larger than the conventionally
assumed value, $f_B=0.1$ (e.g. Haehnelt et al. 1998) in our case.
We have not considered detailed corrections for dust absorption.

In the absence of the strong spectral evolution induced by the accretion
flow transition to low efficiency ADAFs,
the luminosity evolution is generally much slower than in the
case of the spectral evolution. An accelerated evolution occurs in which
luminosities drop dramatically below the critical mass accretion rate,
$\dot{m}_c$. ADAFs appear with their characteristic hard spectra as shown in
Figure 1 (b). In the single population models with no-spectral evolution,
the obtained best-fit QSO evolutionary time scale is shorter than that of the
spectral evolution case.
Figure 4 (a2) and (b2) show the best-fit LFs in various energy bands
in the single population no-spectral evolution model (hereafter NES model)
and no-spectral evolution with multiple populations (NEM model), respectively.
All parameters used in the case of the non-spectral evolution are
given in Table 1. The resulting MF of BH remnants in each model is shown in
Figure 5.  Comparing the SEM model with the NEM model, we find that the
optical LF in NEM model gives a better fit to the observed one
(B band LF of QSO observed by Boyle et al. (1991))
than that of the SEM model even if there are no substantial differences
in the evolving energy-dependent profiles of LFs in this NE- series models.
Despite the physical motivation for the spectral evolution, we conclude
that the NEM model gives the evolving shape of LFs which fits the observed
optical ($4400 \AA$) as well as soft X-ray LFs
even if, in this NE- series models, both of LFs
show the same evolution rates, differently from the observed one.
Wisotzki (2000b) attempts to reconcile the discrepant evolution
rates of optical and X-ray LFs by modifying an optical LF
(i.e. a new determination of the optical K correction for QSOs).
It still remains problematic that the BH number density corresponding to
the observed soft X-ray LF (Miyaji et al. 2000)
is too high to match the number density observed by Boyle et al. (1991).
It appears that our results indicate that the optical and X-ray LFs evolve
separately with different number densities. For instance,
Haiman \& Menou (2000) suggest the possibility that the ratio of
optical to X-ray emission of quasars could evolve with
redshift due to the cosmological structure formation and accompying
evolution of the QSO environment.

The two models, one with the spectral evolution and the other without, have
substantial differences in the hard X-ray luminosity evolution.
Consequently, as previously predicted (Choi et al. 1999b), the observational
LF data in the hard X-ray, which will be obtained by {\it Chandra} X-ray
observatory could play a role in determining the importance of the spectral
evolution. In conclusion, the NEM model is more plausible in accounting for
the observed LF. This result partially supports the possibility that
the QSO population is composed of the multiple short-lived (a few 10$^8$ yrs)
generations of which
the lifetime is short enough to have no significant spectral evolution
over each generation's evolution. This is in agreement with the
observational result that the spectra in the X-ray bands
show no conspicuous variation in
the spectral index with redshift (e.g. Blair et al. 2000).

\section{QSO LFs in the Disk Instability Model}

The observational and theoretical similarities between Galactic X-ray
binaries and AGN in emission mechanisms and spectral behavior
lead us not only to assume the QSO spectrum-luminosity correlation
but also to expect the QSO accretion flows' time-dependent phenomena
similar to those of the Galactic binary systems. Most notably, the
occasional outbursts from black hole binaries in the Galaxy could be
relevant for AGN/QSO emission from scaled-up accretion flows around
supermassive black holes. The enormous differences between Galactic
black hole systems and AGN/QSO should be reflected in time scales of similar
phenomena such as the recurrent outbursts.
If the variability due to the thermal-viscous ionization instabilities
in accretion disks in cataclysmic variables or X-ray transients
does operate in accretion disk in AGN on longer time scales
(Siemiginowska, Czerny, \& Kostyunin 1996 and reference therein,
Mineshige \& Shields 1990), the observed QSO/AGN LFs could be affected
by the recurrent outbursts.
Siemiginowska \& Elvis (1997) assumed that all QSOs are subject to this
variability and calculated the LF of a population of identical sources
with a single mass and an accretion rate while the possible ADAF-like emission
at the low luminosity end of the light curve was ignored.

Here, we consider the disk instability model for QSOs in which QSOs' disk
emission is periodically modulated with recurrent outbursts separated by
quiescence. We apply our simple assumptions on the luminosity evolution of
QSOs to this disk instability model and derive the LFs and MFs for four
sets of evolution models of QSOs; (i) Disk instability induced spectral
evolution model for a single QSO population (DSES model), (ii) Disk
instability induced non-spectral evolution model of a single QSO population
(DNES model), (iii) Disk instability induced spectral evolution model of
multiple QSO populations (DSEM model), and (iv) Disk instability induced
non-spectral evolution model of multiple QSO populations (DNEM model).
The resulting LFs and MFs for these disk instability models are included
in Figure 4 and 5.
In order to obtain the soft X-ray LF in good agreement with
the observed LFs of AGN in the soft X-ray band (Miyaji et al. 2000),
the parameters are optimized and the comoving number density is normalized.
The best-fit parameters determined for various models are listed in Table 1.

We assume that a single QSO undergoes repeated bright and faint phases.
These bright and faint phases correspond to outbursts and quiescent states
respectively. When the spectral evolution model is adopted,
the behavior of low-luminosity part is essentially determined by the ADAF
emission. A number of new parameters are introduced in the disk instability
models. The major model parameters include the amplitude of accretion rate
variation, duty cycle between the active phase and the quiescent phase,
and the number of bursts. We assume that the amplitude of accretion rate
variation is constant throughout the evolution while the maximum and minimum
accretion rates decrease exponentially with the e-folding time scale $t_{evol}$
as defined in Section 2. In the model calculations,
the maximum accretion rate does not exceed the
Eddington accretion rate, $\dot{m}=1$, and the variability is taken to be
periodic in cosmic time. Due to the redshift dependence of time scale,
the variability in the redshift unit depends on the redshift epoch.
The duty cycle, the ratio of the elapsed time of QSO in the active
phase and the quiescent phase, $t_{act}/t_{q}$, essentially characterizes the
outburst time scale in the disk instability model.
There are two important simplifying assumptions we adopt here in order to
make the calculations possible.
First, an identical set of period and amplitude of variabilities
is applied to QSOs regardless of the overall QSO evolution time scale.
Second, because the number of parameters required in a relatively
simple DSES model is already excessive, the inter-dependence among these
parameters is ignored for practical calculational purposes.

\subsection{Single Population Model with Disk Instability}

Figure 6 shows the parameter dependence of the soft X-ray LFs for DSES model.
The thermal instability in the accretion
disk occurs continuously over the cosmological time scale.
See the QSO light curve for a BH mass of $10^8 M_{\odot}$ in panel (a).
The amplitude of accretion rate variation, $\Delta$ log $\dot{m}$ is set at
1 and the accretion rate changes without much gain in mass.
Because the BH gains most of its mass during an early active phase,
the masses of BHs in this model do not grow substantially.
Therefore, it is required that the sample QSOs experiencing
thermal-viscous ionization instabilities in accretion disks
have more massive seed BHs in order to fit the observed LF and MF.
The maximum of accretion rate starts from 1 in Eddington unit at the time
birth and subsequently it falls exponentially below the critical rate,
log $\dot{m}_c =-2$ causing the transition
of accretion flow from thin disk (HS) to ADAF (LS/OS).
A QSO evolving with the larger $\Delta$ log $\dot{m}$ experiences the ADAF phase
earlier and longer than one with the smaller amplitude.
The ratio of $t_{act}/t_{q}=0.25$ and the number of outbursts of $10^3$ mean
that QSO spends about $20\%$ in the active phase and about $80\%$
in the quiescent phase and the time scale for a burst in the active phase,
$t_{act}$, is of the order of $10^6$ yr, respectively.

From the light curve in panel (a) in Figure 6,
we see the dominant redshift dependence of time scale which shows that in the
redshift plot the recurrence time scale shortens toward low redshifts.
Panel (b1) gives the LFs which fit the observed ones in
this DSES model. In Figure 4 (a3) and (a4),
the LFs in various energy bands for DSES and DNES models are displayed.
In these disk instability models,
the number density of bright QSOs does not remain
constant and it decreases just as in the SEM and NEM models.
This is simply because the dim QSOs in quiescent phase effectively disappear
in number counting.
Based on the comparisons between SES and DSES models and between NES and DNES
models, we see that there are little differences between them in the overall
shape of LFs, which in part implies that the adopted disk instability model
(see the best-fit parameters in Table 1) may not be sophisticated enough
to catch some potential subtle differences. Within the accuracy of the current
models, it is plausible that the QSO accretion disks may indeed undergo
the thermal instability and that the observed LFs may well be accounted for
by the disk instability model. This possibility is not directly testable in
individual QSOs because the variability time scales expected for AGN/QSO in
our model are simply too long (e.g. $10^5 \sim 10^6$ yr) and is not directly
observable.

From panel (b1) {\it vs.} (b4) and panel (b2) {\it vs.} (b3) in Figure 6,
we see the effects of the amplitude of accretion rate variation
on the LFs in DSES model. The amplitude at the level of
$\Delta$ log $\dot{m}=4$ allows the QSO accretion disk
to undergo recurrent spectral changes between the HS and the LS/OS.
Starting from its birth, a QSO with a time-varying accretion disk spends
most ($\sim 80\%$) of its evolution in the quiescent phase. The quiescent
phase is comprised of LS and OS as the mass accretion rate fluctuates.
A large number of QSOs remain in the very dim luminosity state while
dramatically increasing the undetected low luminosity galactic nuclei.
This increase in the number density which is normalized to match
the comoving number density of Miyaji et al. (2000) depends on the
amplitude of the mass accretion rate fluctuations. For instance,
the derived comoving number density of QSOs (both observed and dormant)
in the case of $\Delta$ log $\dot{m}=4$ is roughly 10 times higher than
that in the case of $\Delta$ log $\dot{m}=1$.
From panel (b1) {\it vs.} (b2) and panel (b3) {\it vs.} (b4),
we show the dependence of the LF on the duty cycle in the DSES model, and
from from panel (b1) {\it vs.} (b5), the dependence on the burst frequency.
The number of bursts $10^4$ means that $t_{act}$ for a burst
is of the order of $10^5$ yr. As the duty cycle shortens,
the time elapsed in the quiescent phase of QSO increases and therefore,
the more QSOs tend to stay in the low luminosity part of LF at any given
epoch.  The resulting LFs in the panels, however, show that
the number of bursts and the duration of the duty cycle do not affect
the evolving shape of the LF significantly in contrast to the effects of the
amplitude of accretion rate variation.

\subsection{The Multiple Population Model with Disk Instability}

In Figure 4, panel (b3) and (b4) show the best-fit LFs in
various energy bands for DSEM and DNEM models.
MFs of remnant BHs corresponding to these LFs appear in Figure 5.
Among the multiple population models,
there is little difference in the evolving shape of LFs.
It is remarkable that even the evolving shape of LF of the DSEM model
with $\Delta$ log $\dot{m}=4$ is not conspicuously different from those
of the other multiple population models. This is interesting because
the derived comoving number density for this very large amplitude case
is substantially bigger (by a factor $\sim 10$) than those of the other models.
This example demonstrates that the observed LF alone cannot rule out a
large number of dormant and hence undetected QSO/AGN at any given epoch.
However, this large comoving number density value is still short of
the required level in the MF. We conclude that the disk instability
model cannot simultaneously account for the LF and MF.

\section{DISCUSSIONS}

We have explored diverse types of possible phenomenological scenarios
for the cosmological evolution of QSOs. Based on the basic paradigm in which
QSO/AGN are powered by accreting supermassive black holes
(Frank et al. 1992), we have considered
various spectral states correlated with the luminosity level
(Choi et al. 1999b) and constructed
the analytical LFs in the various energy bands. Using the MF
of remnant BHs in nearby galactic nuclei as an additional constraint, we have
examined whether there exists a model simultaneously satisfying both LF and
MF. The best and hence most promising scenario along with the best fit
parameters points to the following formation and accretion history for
supermassive BHs and QSOs.

\begin{enumerate}
\item In the multiple population models, due to the rapid decrease of $\dot M$
      caused by the short evolutionary time scale and to the redshift dependence
      of the evolutionary time scale, the number density of bright QSOs has to
      evolve strongly over the cosmological time.
      According to the derived soft X-ray LFs in a good agreement
      with observed LFs (Miyaji et al. 2000),
      the QSOs with less massive seed BHs must have been born more abundantly
      and/or refueled more efficiently at lower redshifts.

\item While the space number density obtained from the soft X-ray LF of AGN
      by (Miyaji et al. 2000) is about only 3 times higher than that
      observed by Boyle et al. (1991),
      the resulting comoving space densities in the BH remnants inferred from
      the derived best-fit soft X-ray LFs in all QSO evolution models,
      which we have tested, are smaller than those estimated for the putative
      BHs in nearby galaxies by a factor of about 10 to 100.
      No models we have explored can derive the LF and MF satisfying both
      of the two observational constraints well enough. This could imply that
      it is highly unlikely that all galaxies contain MDOs which are actually
      supermassive BH remnants of QSOs. Alternatively, there could be a
      population of highly obscured AGN which harbor growing BHs
      (Fabian 1999, Hasinger 2000).

\item There is little difference in evolving shapes of
      soft X-ray LFs among all the models we considered. This is inevitable
      given the fact that our models have at least been required to fit
      the observed LFs of QSO/AGN in the soft X-ray bands (Miyaji et al.
      2000). We, however, find substantial differences in optical ($4400 \AA$)
      LFs and hard X-ray LFs, which is a natural consequence of the assumed
      spectrum-luminosity correlation (Choi et al. 1999a, 1999b, 2000).
      In the case of the spectral evolution, the optical LF evolves more
      rapidly than the X-ray LF especially at lower redshifts or when QSOs
      become faint. There is a great difference among the evolving features
      of the hard X-ray LFs in various models. This is one of the strongest
      outcomes in the spectral evolution models.
      Expected observational data by hard X-ray missions such as
      the {\it Chandra} X-ray observatory could play a role
      in providing a much needed, additional observational constraint.
      Such a constraint will shed light on which QSO models are more plausible
      and provide a statistical test on whether QSOs have experienced the
      spectral transition over the cosmological time scale.
      Such a transition is obviously untestable in individual galaxies
      due to the prohibitively long evolution and transition time scales.

\item The multiple population model with no spectral evolution gives the
      evolving shape of LFs which fit the observed B band LF of QSO derived
      by Boyle et al. (1991) in the optical ($4400 \AA$) band as well
      as the soft X-ray LF. It is by far the best model among the various
      evolution models we have tested. This result supports the possibility
      that the QSO population is composed of many short-lived (a few 10$^8$ yr)
      generations
      while in each short generation QSOs do not experience any
      significant spectral evolution.
\end{enumerate}

This work was supported by a KRF grant No. 1999-001-D00365.
JY was also supported by the MOST through the National R \& D
program (99-N6-01-01-A-06) for women's universities. IY wishes to thank
Anna I. Yi and Fojyik Yi for numerous discussions and helpful suggestions and
Ethan Vishniac for hospitality while this work was in progress.

\clearpage
\vfill\eject

\begin{table}
\caption
{Best-Fit Parameters for LFs in Various Models Considered}
\begin{center}
\begin{tabular}{lc}
\hline\hline
 Model$^{\rm a}$ &  Parameters$^{\rm b}$\\
\hline
\\
SES  & $t_{evol}=0.50t_{age}$, $z_i=4$($\sigma_z=0.5$),
       $\dot m_i=1$($\sigma_{\dot m}=0.1$)\\
NES  & $t_{evol}=0.25t_{age}$, $z_i=4$($\sigma_z=0.5$),
       $\dot m_i=1$($\sigma_{\dot m}=0.1$)\\
DSES & $t_{evol}=0.40t_{age}$, $z_i=4$($\sigma_z=0.5$),
       $\dot m_i=1$($\sigma_{\dot m}=0.1$);\\
     & $\Delta$log$\dot m=1, t_{act}/t_q =0.25, 10^3$ bursts\\
DNES & $t_{evol}=0.25t_{age}$, $z_i=4$($\sigma_z=0.5$),
       $\dot m_i=1$($\sigma_{\dot m}=0.1$);\\
     & $\Delta$log$\dot m=1, t_{act}/t_q =0.25, 10^3$ bursts\\
\hline
SEM  & $t_{evol}=0.03t_{age}$, $\dot m_i(z)=1$($\sigma_{\dot m}=0.01$);\\
     & $g=50$, $(z_o)_i=4.3, (z_f)_i=0.18$($\sigma_z=0.001$)\\
NEM  & $t_{evol}=0.05t_{age}$, $\dot m_i(z)=1$($\sigma_{\dot m}=0.01$);\\
     & $g=50$, $(z_o)_i=4.3, (z_f)_i=0.18$($\sigma_z=0.001$)\\
DSEM & $t_{evol}=0.03t_{age}$, $\dot m_i(z)=1$($\sigma_{\dot m}=0.01$);\\
     & $g=50$, $(z_o)_i=4.5, (z_f)_i=0.18$($\sigma_z=0.001$);\\
     & $\Delta$log$\dot m=4, t_{act}/t_q =0.25, 10^3$ bursts\\
DNEM & $t_{evol}=0.03t_{age}$, $\dot m_i(z)=1$($\sigma_{\dot m}=0.01$);\\
     & $g=50$, $(z_o)_i=4.5, (z_f)_i=0.18$($\sigma_z=0.001$);\\
     & $\Delta$log$\dot m=1, t_{act}/t_q =0.25, 10^3$ bursts\\
\hline
\end{tabular}
\end{center}
$^{\rm a}$ D-: Disk instability,
SE-: spectral evolution, NE-: No-spectral evolution,
S-: Single QSO population, M-: Multiple QSO populations.
$^{\rm b}$ $H_o =50$km s$^{-1}$ Mpc$^{-1}$, $q_o =0.5$, $\Omega _{\Lambda}=0$,
$L_{bol}$: bolometric luminosity,
$L_{HXR+SXR}=0.1L_{bol}$ in SE- series models,
$L_{HXR}=0.04L_{bol}$, $L_{SXR}=0.06L_{bol}$, and
$L_{4400\AA}=0.43L_{bol}$ in NE- series models. The functional forms in the
multiple population models correspond to
$N(z)=N_{o}[(1+z)/(1+z_{o})]^{-1}$, where $N_{o}$ is the number of QOSs
belonging to the first generation at the epoch, $z=z_{o}$. $z_{f}$
represents the epoch at which the final generation is born.
\end{table}

\clearpage
\noindent
\vfill\eject
\clearpage

\begin{figure}
\caption{
(a) Spectral energy distributions of a QSO with a BH mass of
$10^8 M_\odot$.
From top to bottom, the decreasing $\dot m$ causes the spectral and
luminosity evolution. VHS, HS, LS$+$OS denote
the very high state, high state, low state including the off
state, respectively. (b) A luminosity evolution of a QSO
for $t_{evol} \simeq 6.4 \times 10^9$ yr with an initial BH mass of
$10^8 M_\odot$, $\dot m =1$. During accretion, different energy bands show
different evolutions. At hard X-ray energies, the characteristic transition is
identified below $z\sim 1$, at which the luminosity in each band
decreases more steeply than the one based on the bolometric luminosity
evolution (a dash-dotted line).
}
\end{figure}

\begin{figure}
\caption{
Panels (a1)-(a5) show the dependence of the soft X-ray LF of QSOs
on several major parameters required in the multiple population model
with the spectral evolution correlated with luminosity (i.e. SEM model).
Panel (a1) is the best fit in SEM model.
Panel (a1) and (a2) show the differences
due to the functional form of the number density evolution.
Panels (a1) and (a3) show the effects of the initial mass distribution
of seed BHs at each generation on LF.
Panels (a1), (a4), (a5) show the effects
of superposing multiple generations and varying evolution time scales
on LF.
Panel (c1) shows the mass distribution of BHs at various redshifts
in the multiple population model corresponding to the best fit
model panel (a1).
The solid, dash-dotted (without symbols), and dashed
lines in panel (c1) represent the MF of BH remnants corresponding to the LFs
in panels (a1), (a4), and (a5), respectively,
and the solid lines with the cross symbol in panels (c1) and (c2)
represent the estimated MF from the  radio data of E/SO galaxies
in Salucci et al. (1999).
The comoving number density is normalized to match that of
Miyaji et al. (2000). Panel (c2) shows the MF in the single population model.
}
\end{figure}

\begin{figure}
\caption{
The soft X-ray
LF in panel (a1) is responsible for the MF
of BH remnants (dashed lines in panel (c)) of
multiple QSO population without the spectral evolution
(i.e. NEM model) which is the best-fit of
the MF (solid line with cross symbols in panel (c)) estimated from the radio LF.
The parameters required here are given in panel (a1).
LF in panel (a2) is the best-fit NEM
model for the observed soft X-ray LF of (b)
and the dash-dotted line in panel (c) represents the MF corresponding to
this LF. The comoving number density is normalized to match that of
Miyaji et al. (2000) in panel (b).
}
\end{figure}

\clearpage
\noindent
\vfill\eject
\clearpage
\begin{figure}
\caption{
The redshift evolutions of the LF for single (S-)
and multiple (M-) population models (panels (a) and (b), respectively).
The LF evolution has been obtained under the set of four assumptions
on the luminosity evolution of a QSO; spectral evolution (SE-), no-
spectral evolution (NE-), and disk instability induced evolutions in
both types of spectral evolution (D-).
By demanding that we obtain the soft X-ray LF in a good agreement with
the observed soft X-ray LFs (Miyaji et al. 2000),
we fix the parameters and normalize the comoving number density.
The best-fit parameters used in various models are shown
in the Table 1.
}
\end{figure}

\begin{figure}
\caption{
MFs of BH remnants corresponding to the best-fitted LFs shown in
Figure 3. The comoving number densities of derived QSO remnants
are short of the required numbers for the MF estimated from the radio LF
(Salucci et al. 1999).
}
\end{figure}

\begin{figure}
\caption{
Panel (a) shows the light curve of a QSO
with an initial BH mass $10^8 M_\odot$ in the disk instability induced
evolution model.
Panels (b1)-(b5) show the dependence of the soft X-ray LF on several
parameters in a model with a single long-lived QSO population.
The QSO population undergoes the long-term luminosity fluctuations
and spectral evolution (i.e. DSES model) with the main parameters being
the duty cycle between the active and quiescent phases, $t_{act}/t_q$,
the amplitude of accretion rate variation, $\Delta$log$\dot m$,
and the number of bursts. Panel (b1) shows the best fit in this model.
The comoving number density is normalized to match that of
Miyaji et al. (2000) in panel (c).
}
\end{figure}


\begin{references}
\reference{A} Boyle, B. J., Jones, L. R., Shanks, T., Marano, B., Zitelli, V.,
\& Zamorani, G. 1991, in The Space Distribution of Quasars, ASP Conference
Series 21, ed. D. Crampton, Astron. Soc. Pacif., San Francisco, p.191
\reference{A} Boyle, B. J., Shanks, T., Croom, S. M., Smith, R.
J., Miller, L., Loaring, N., \& Heymans, C. 2000, preprint
(astro-ph/0005368)
\reference{A} Blair, A. J., Stewart, G. C., Georgantopoulos, I.,
Boyle, B. J., Griffiths, R. E., Shanks, T., \& Almaini, O. 2000,
MNRAS, 314, 238
\reference{A} Caditz, D., Petrosian, V., \& Wandel, A. 1991, ApJ, 372, L63
\reference{A} Choi, Y., Yang, J., \& Yi, I. 1999a, J. Korean Phys. Soc., 34,
L199
\reference{A} Choi, Y., Yang, J., \& Yi, I. 1999b, ApJ, 518, L77
\reference{A} Choi, Y., Yang, J., \& Yi, I. 2000, Nuovo Cimento B.
in press
\reference{A} Comastri, A., Setti G., \& Zamorani, G., 1995, A\&A, 296, 1
\reference{A} Di Matteo, T., Esin, A., Fabian, A. C., \& Narayan, R. 1999,
MNRAS, 305, L1
\reference{A} Esin, A. A., McClintock, J. E., \& Narayan, R. 1997, ApJ, 489,
865
\reference{A} Fabian, A. C. 1999, MNRAS, 308, 39
\reference{A} Fabian, A. C. \& Canizares, C. R. 1988, Nature, 333, 829
\reference{A} Fabian, A. C. \& Rees, M. J. 1995, MNRAS, 277, L55
\reference{A} Franceschini, A., Vercellon, S. \& Fabian, A. C.
1998, MNRAS, 297, 817
\reference{A} Franceschini, A., Hasinger, G., Miyaji, T. \& Malquori, D.
1999, MNRAS, 310, L5
\reference{A} Frank, J., King, A. R., \& Raine, D. 1992, Accretion Power in
Astrophysics (Cambridge: Cambridge Univ. Press)
\reference{A} Fukugita, M. \& Turner, E. L. 1996, ApJ, 460, L81
\reference{A} Hartwick, F. D. A. \& Shade, D. 1990, ARA\&A, 28, 437
\reference{A} Haehnelt, M. G. \& Rees, M. J. 1993, MNRAS, 263, 168
\reference{A} Haehnelt, M. G., Natarajan, P. \& Rees, M. J. 1998, MNRAS,
300, 817
\reference{A} Haiman, Z. \& Hui, L. 2000, preprint
(astro-ph/0002190)
\reference{A} Haiman, Z. \& Menou, K. 2000, ApJ, 531, 42
\reference{A} Hasinger, G. 1998, Astron. Nachr. 319, 37
\reference{A} Hasinger, G. 2000, preprint (astro-ph/0001360)
\reference{A} Hatziminaoglou, E., Waerbeke, L. V., \& Mathez, G. 1998, A\&A,
335, 797
\reference{A} Kauffman, G. \& Haehnelt, M. 2000, MNRAS, 311, 576
\reference{A} Madau, P., Chisellini, G., Fabian, A. C., 1994, MNRAS, 270, L17
\reference{A} Magorrian, J. et al. 1998, AJ, 115, 2285
\reference{A} Mahadevan, R. 1997, ApJ, 477, 585
\reference{A} Mathez, G. 1976, A\&A, 53, 15
\reference{A} Mathez, G. et al. 1996, A\&A, 316, 19
\reference{A} Mineshige, S. \& Shields, G. A. 1990, ApJ, 351, 47
\reference{A} Miyaji, T., Hasinger, G., \& Schmidt, M. 2000, A\&A, 353, 25
\reference{A} Monaco, P., Salucci, P., \& Danese, L. 2000, MNRAS, 311, 279
\reference{A} Narayan, R., Mahadevan, R., \& Quataert, E. 1998, in Theory of
Black Hole Accretion Disks, eds. Marek A. Abramowicz, G. Bjornsson,
\& James E. Pringle, Cambridge, Cambridge Univ. Press, 148p
\reference{A} Narayan, R. \& Yi, I. 1995, ApJ, 452, 710
\reference{A} Nulsen, P. E. J. \& Fabian, A. C. 2000, MNRAS, 311, 346
\reference{A} Peebles, P. J. E. 1993, Principles of Physical Cosmology
(Princeton: Princeton Univ. Press)
\reference{A} Peterson, B. M. 1997, An Introduction to Active Galactic Nuclei
(Cambridge: Cambridge Univ. Press)
\reference{A} Phinney, E. S. 1997, presented at IAU Symposium No. 186 Galaxy
Interactions at Low and High Redshift, Kyoto, Japan, 94p
\reference{A} Rutledge, R. E. et al. 1999, ApJS, 124, 265
\reference{A} Siemiginowska, A., Czerny, B., \& Kostyunin, V. 1996, ApJ, 458,
491
\reference{A} Siemiginowska, A. \& Elvis, M. 1997, ApJ, 482, L9
\reference{A} Small, T. A. \& Blandford, R. D. 1992, MNRAS, 259, 725
\reference{A} Salucci, P., Szuszkiewicz, E., Monaco, P., \& Danese, L. 1999,
MNRAS, 307, 637 (Erratum in 311, 448)
\reference{A} Wandel, A. 1999, ApJ, 519, L39
\reference{A} Weedman, D. W. 1986, Quasar Astronomy (Cambridge: Cambridge Univ.
Press)
\reference{A} Wisotzky, L. 2000a, A\&A, 353, 853
\reference{A} Wisotzky, L. 2000b, A\&A, 353, 861
\reference{A} Yi, I. 1996, ApJ, 473, 645
\reference{A} Yi, I. \& Boughn, S. P. 1998, ApJ, 499, 198
\reference{A} Yi, I. \& Boughn, S. P. 1999, ApJ, 515, 576
\end{references}
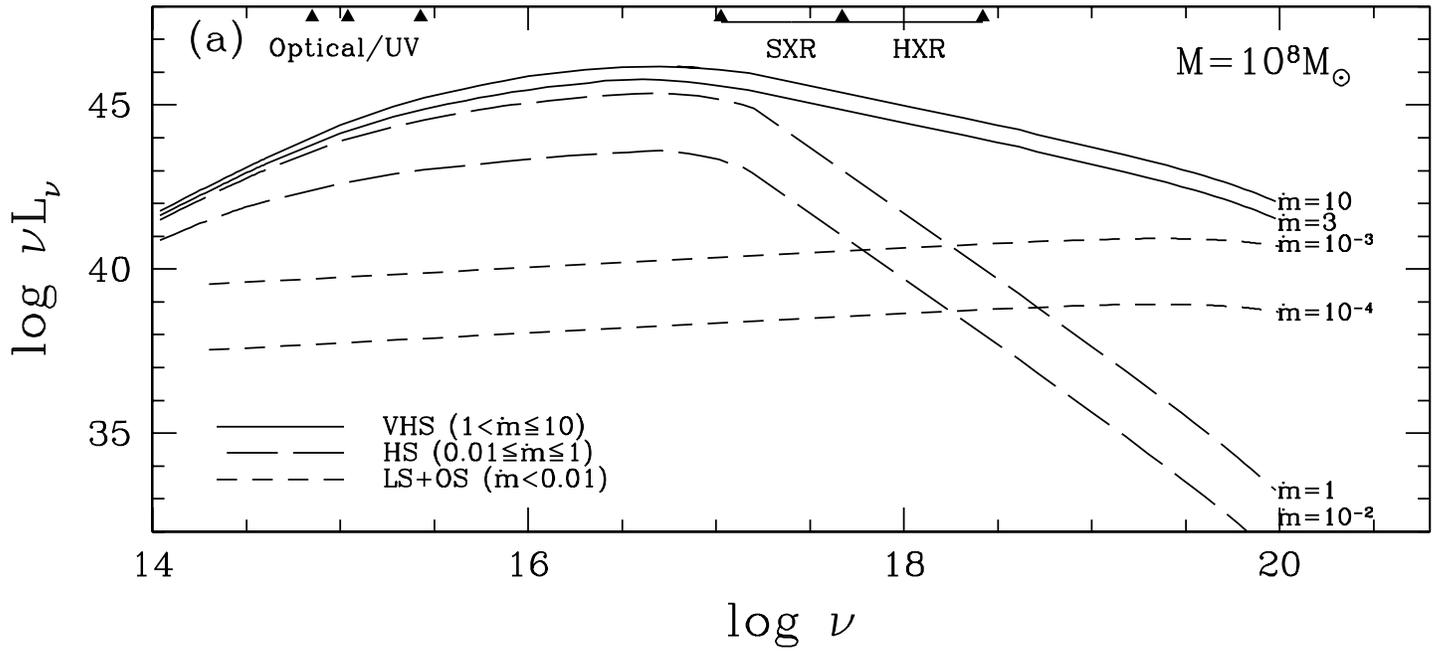
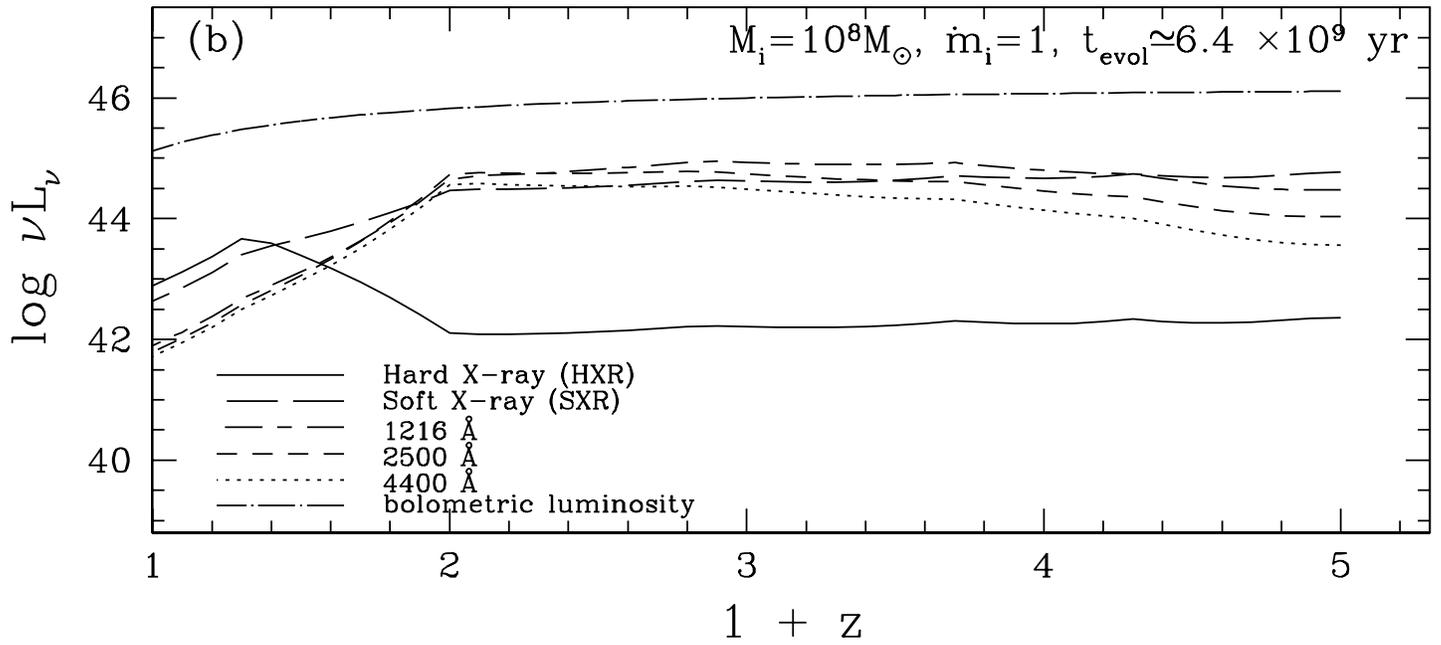
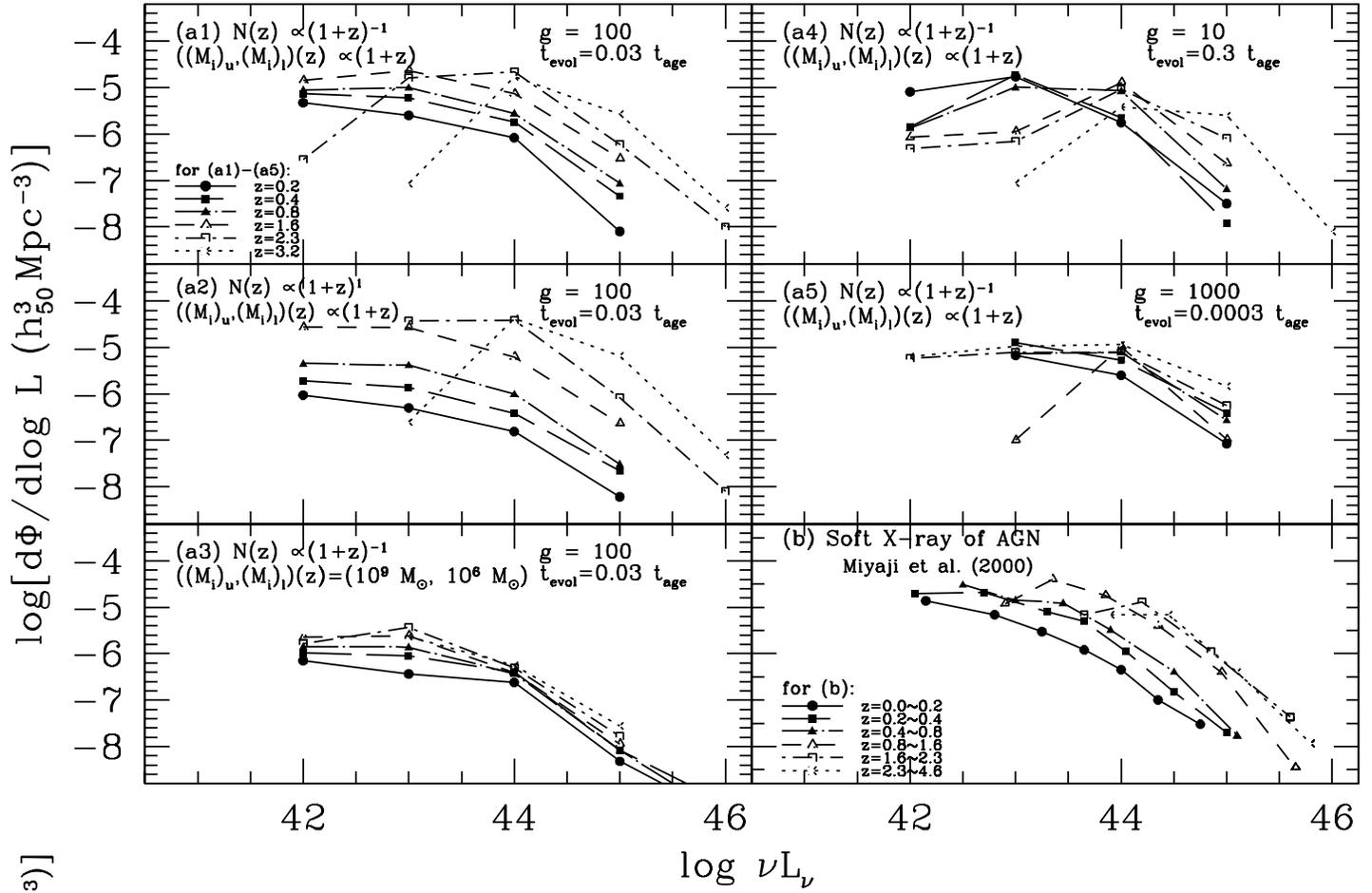
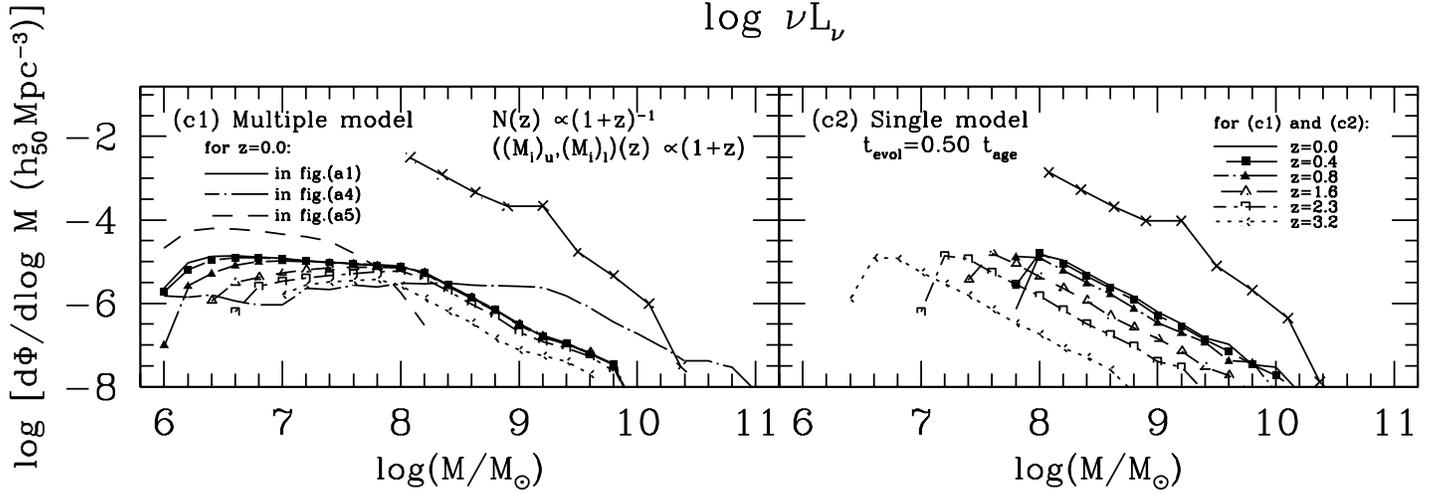
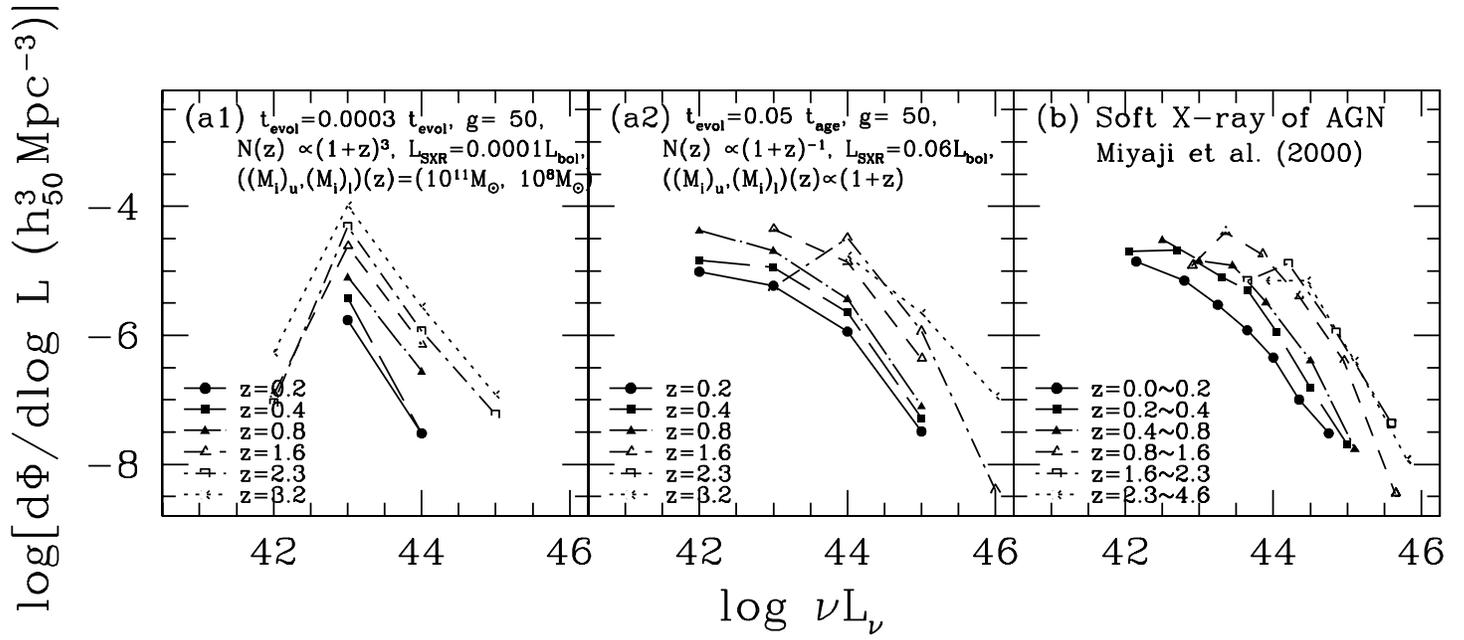
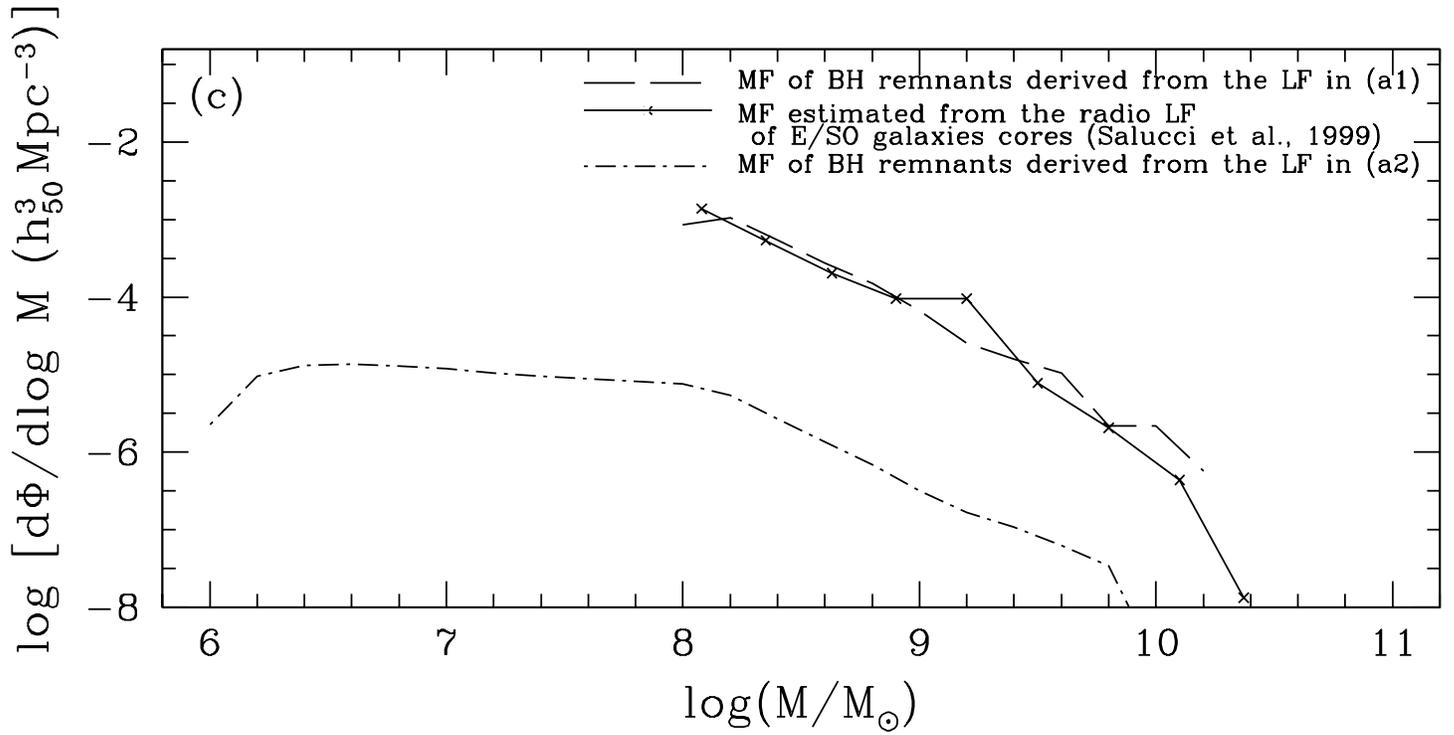
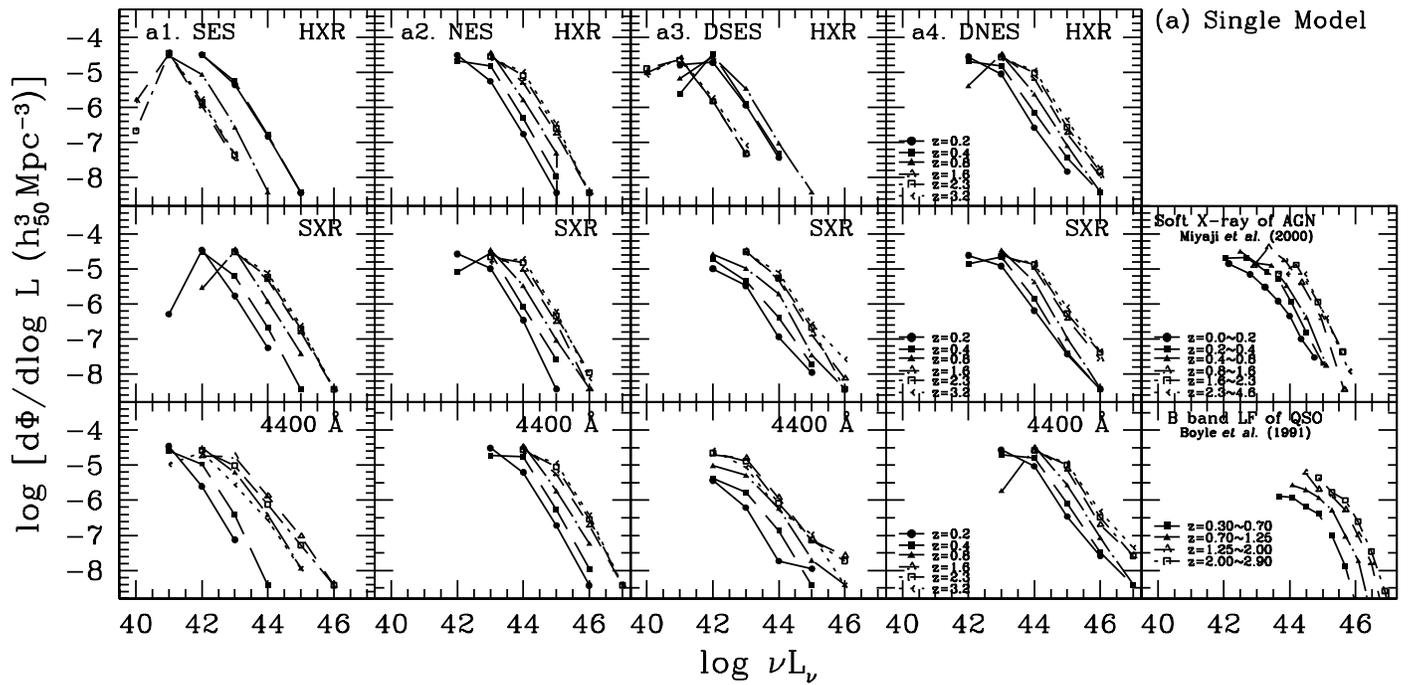
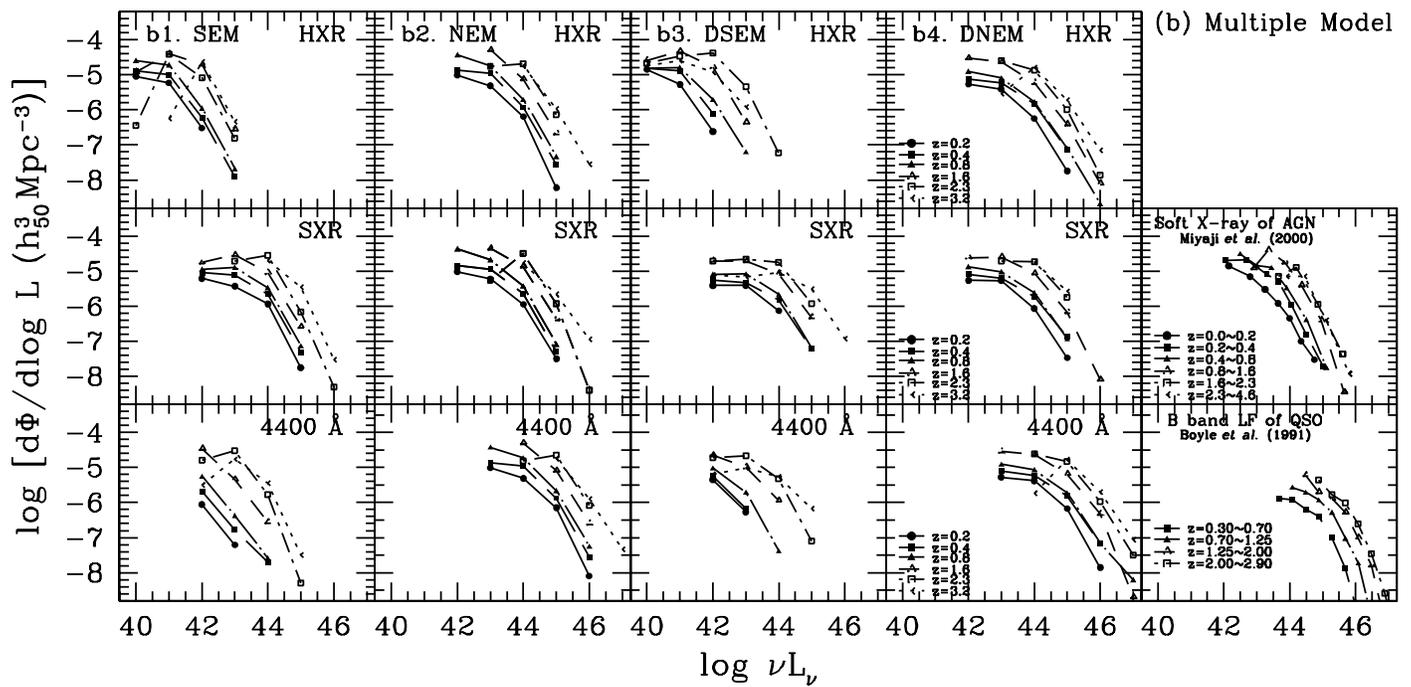
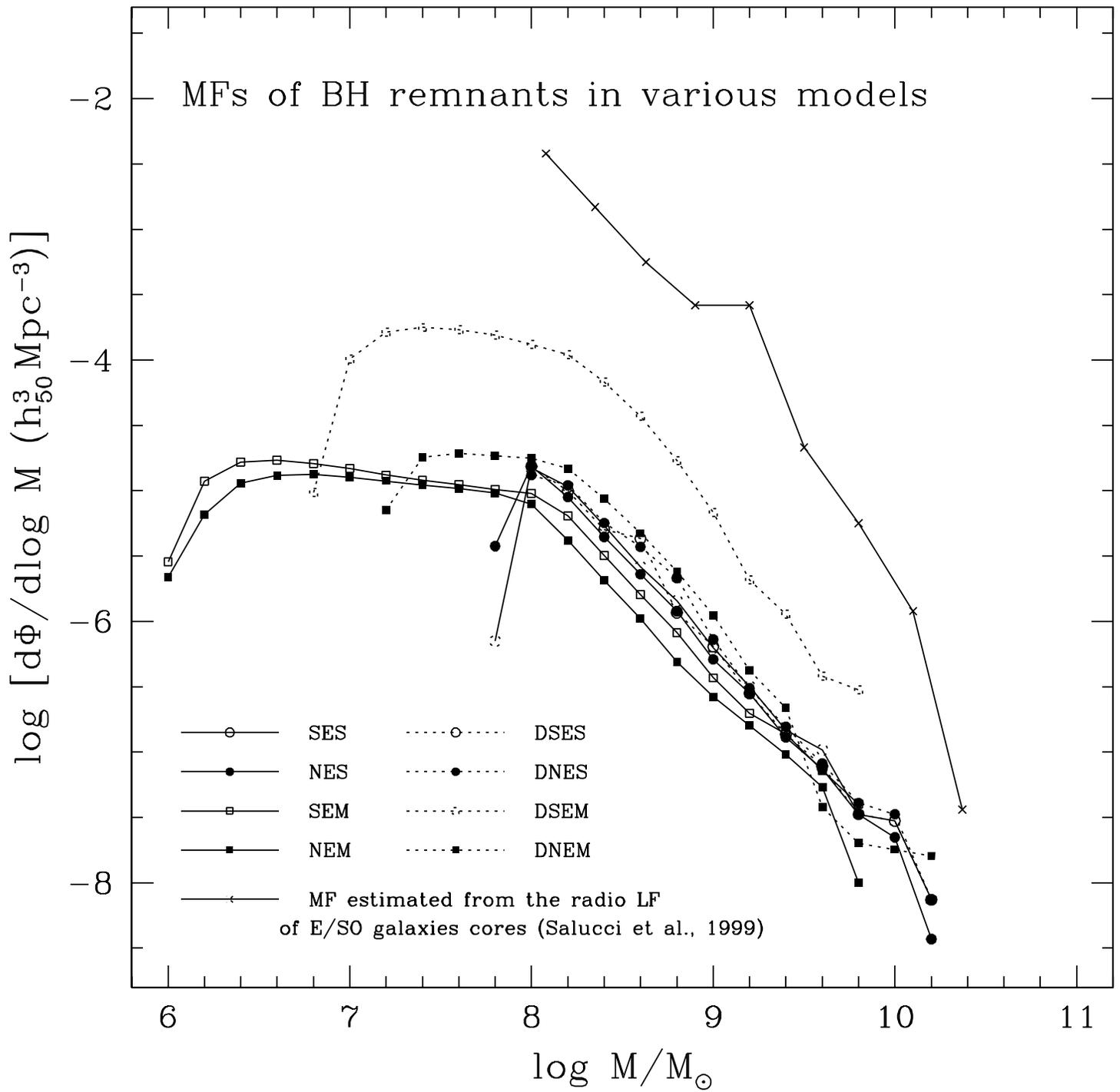
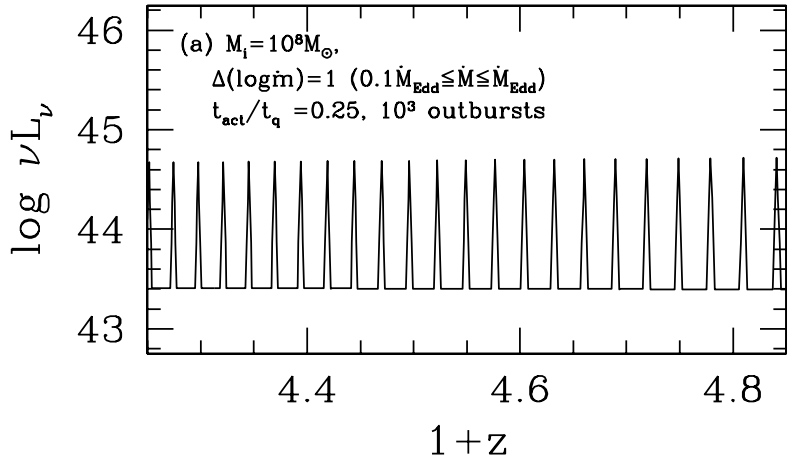
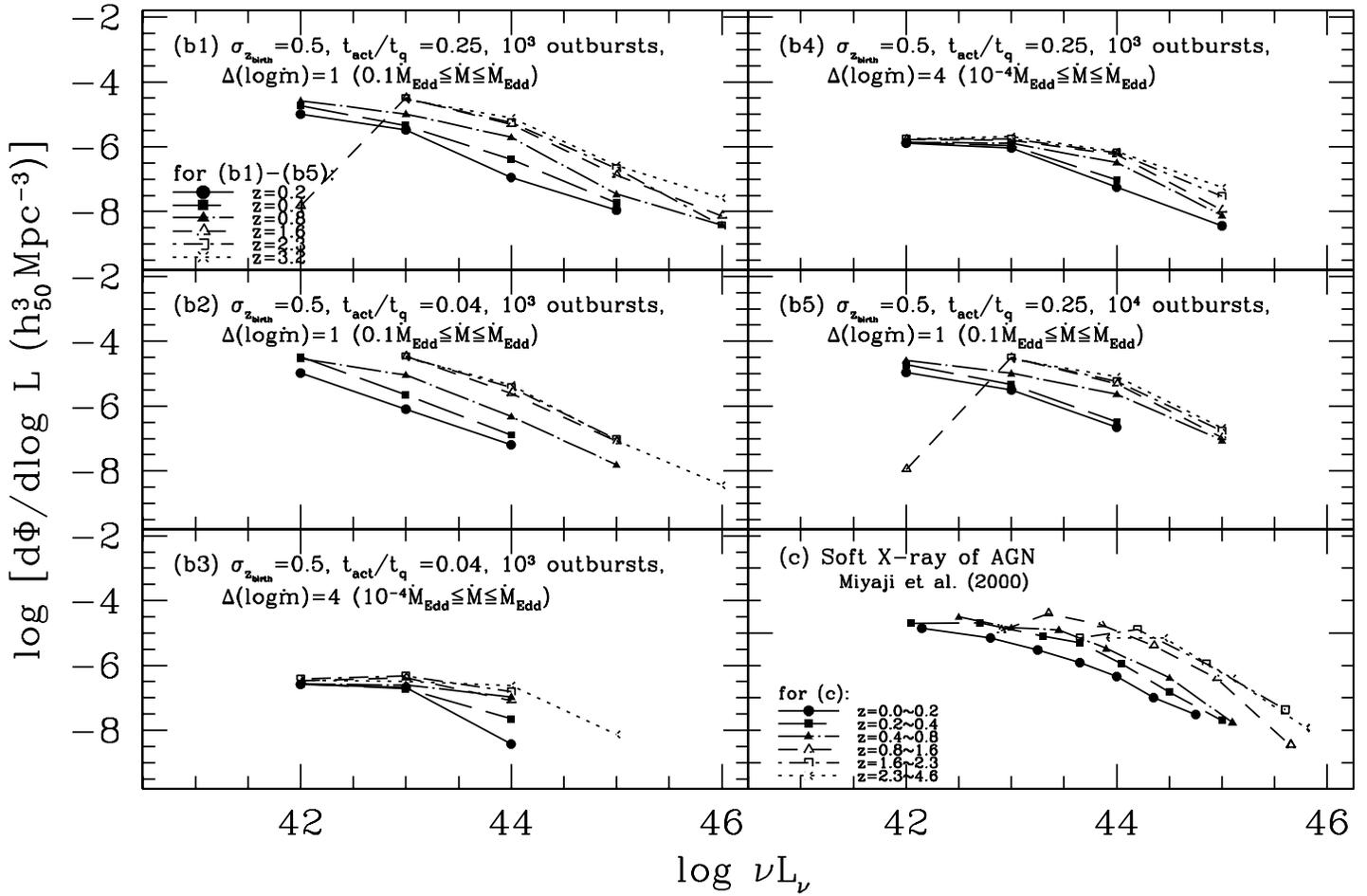
\end{document}